\documentclass[journal]{IEEEtran}
\usepackage{epsf}
\begin{document}
%
% paper title
\title{Probing the Phase Diagram of 
Bi$_{2}$Sr$_{2}$CaCu$_{2}$O$_{8+\delta}$ with 
Tunneling Spectroscopy}

\author{L. Ozyuzer, J. F. Zasadzinski, K. E. Gray, D. G. Hinks and N. Miyakawa }% <-this % stops a space
\thanks{Manuscript received August 6, 2002; revised December 9, 2002.
        This research is supported by the U.S.\ Department of Energy, Basic 
Energy Sciences-Materials Sciences, under contract number 
W-31-109-ENG-38 and TUBITAK (Scientific and Technical Research 
Council of Turkey) project number TBAG-2031.  L.O.  acknowledges 
support from Young Investigators Award of Turkish Academy of Sciences.
N.M.  acknowledges support from Grant-in-Aid for Encouragement of Young 
Scientists from the Ministry of Education, Science and Culture, Japan.}% <-this % stops a space
\thanks{L. Ozyuzer is with the Izmir Institute of Technology, Department of Physics, Gulbahce Campus, Urla, TR-35437 Izmir, Turkey; phone: +90 232 4987518; fax: +90 232 4987518; e-mail: ozyuzer@likya.iyte.edu.tr}
\thanks{J. F. Zasadzinski is with the Illinois Institute of Technology, Physics Division, Illinois, USA}
\thanks{K. E. Gray and D. G. Hinks are with the Argonne National Laboratory, Materials Science Division, Illinois, USA}
\thanks{N. Miyakawa is with the Tokyo University of Science, Suwa, Japan}

% make the title area
\maketitle

\begin{abstract}
Tunneling measurements are performed on Ca-rich single crystals of 
Bi$_{2}$Sr$_{2}$CaCu$_{2}$O$_{8+\delta}$ (Bi2212), with various oxygen 
doping levels, using a novel point contact method.  At 4.2 K, SIN and 
SIS tunnel junctions are obtained with well-defined quasiparticle 
peaks, robust dip and hump features and in some cases Josephson 
currents.  The doping dependence of tunneling conductances of Ca-rich 
Bi2212 are analyzed and compared to stoichiometric Bi2212.  A 
similar profile of energy gap vs.\  doping concentration is found 
although the Ca-rich samples have a slighly smaller optimum 
T$_c$ and therefore smaller gap values for any doping level.  The 
evolution of tunneling conductance peak height to background ratios 
with hole concentration are compared.  For a given doping level, the 
Ca-rich spectra showed more broadened features compared to the 
stoichiometric counterparts, most likely due to increased disorder 
from the excess Ca.  Comparison of the dip and hump features has 
provided some potential insights into their origins.
\end{abstract}


\section{Introduction}

% Here goes your work. It should be OK to use \PARstart
% and looks nice too. However, some conferences don't use
% the large first letter.
\PARstart{T}{he} peculiar doping dependencies 
of the superconducting properties of Bi$_{2}$Sr$_{2}$CaCu$_{2}$O$_{8+\delta}$ 
(Bi2212) has stimulated interest in similar measurements on other high temperature superconductors 
(HTSs).  It is necessary to establish 
what properties are universal, not only for all high-$T_{c}$ cuprates, 
but also for various non-stoichiometric compositions of a particular 
HTS.  It is also useful to consider different doping routes other than 
oxygen content.  The doping dependence of the superconducting energy gap, 
$\Delta$, is now well established for 
Bi2212 from tunneling 
\cite{Miyakawa,Renner} and ARPES \cite{Campuzano} studies.  The magnitude of 
the energy gap scales with $T_{c}$ 
in the overdoped regime of Bi2212 indicating conventional behavior, 
however, in the underdoped phase, the energy gap magnitude increases 
with decreasing hole concentration even as $T_{c}$ drops to 70 K from 
its optimal value of 95 K.  Furthermore, the robust spectral features, 
such as dip and hump, also persist over a wide oxygen doping range of 
Bi2212, which offers a link between these features and the 
superconductivity mechanism \cite{Zasadzinski}. The tunneling results 
on Bi2212 also support a novel contention that the doping dependence of the 
energy gap and the pseudogap temperature $T^{*}$ 
closely follow each other suggesting that the pseudogap state is due 
to precursor superconductivity \cite{Oda}. 

In the present paper, we investigated the tunneling conductances 
in Bi$_{2.1}$Sr$_{1.4}$Ca$_{1.5}$Cu$_{2}$O$_{8+\delta}$ (Ca-rich 
Bi2212) by means of superconductor-insulator-normal metal (SIN) and 
superconductor-insulator-superconductor (SIS) junctions, and compared 
the results with stoichiometric Bi2212, to obtain information on the 
influence of the Sr/Ca ratio on the quasiparticle density of states 
(DOS).  It has been previously found 
that the distances between CuO$_{2}$ layers and between SrO and 
Bi$_{2}$O$_{2}$ layers along c-axis become shorter as  Bi2212 
becomes Ca rich.  However, the change of the oxygen concentration does 
not vary any structural parameters \cite{Iwamatsu}. We would expect 
that the shortening c-axis length increases the coupling between 
CuO$_{2}$ planes, giving a tendency for $T_{c}$ and the interplane 
Josephson current to increase, as in the case of uniaxial 
compression \cite{Tajima}. On the other hand, the excess Ca in the 
system resides on Sr sites and creates disorder on the SrO planes next 
to the CuO$_{2}$ planes resulting in the decrease of maximum 
$T_{c}$ \cite{Tokita}. The two combined effects lead to a system which 
is sufficiently different from stoichiometric Bi2212 to allow a 
comparative study.

Magnetic susceptibility studies on Ca-rich Bi2212 
indicated that $T_{max}$, a broad peak in the $T$-dependent curve of 
magnetic susceptibility occurring above $T_{c}$ and $T^{*}$, does not 
depend on the Sr/Ca ratio \cite{Tokita}. The value $k_{B}T_{max}$ is 
considered as a characteristic energy for the effective 
antiferromagnetic (AF) interaction of Cu-spins and the decrease of 
magnetic susceptibility arises from the gradual development of AF 
spin fluctuations.  It is found that 2$\Delta$$\sim$2$k_{B}T_{max}$ for 
stoichiometric 
Bi2212 over a wide doping range.  It might be expected therefore that 
the magnitude of the energy gap in Ca-rich Bi2212 would not change much 
for a given doping level \cite{odareview}. The present study sheds 
light on these issues.

\section{Experiment}

We grew high quality single crystals of Ca-rich Bi2212 using the 
floating zone process.  As grown crystals have a $T_{c}$ of 72-74 K 
with superconducting transition width, $\Delta$$T$, less than 2 K as 
measured by ac magnetization.  We add extra Bi to the system because 
of the necessity of excess Bi content for obtaining single phase 
Bi2212 \cite{Majewski}. A maximum $T_{c}$, $T_{c,max}$, of 81 K is 
obtained by annealing in oxygen indicating that $T_{c,max}$ of 
stoichiometric Bi2212 drops 14 K with excess Ca 
in the crystal. For this study, the doping range of as grown 
crystals extended from overdoped ($T_{c}$=69 K, $\Delta$$T$$<$0.5 K and 
$T_{c}$=77 K, $\Delta$$T$$<$0.3 K), optimally doped ($T_{c}$=81 K, 
$\Delta$$T$$<$1 K) to the underdoped region ($T_{c}$=74 and 72 K 
$\Delta$$T$$<$2 K and $T_{c}$=62 K $\Delta$$T$$<$1.2 K), where overdoped 
samples were obtained by annealing in oxygen and underdoped samples 
by annealing in Ar gas flow.  Both SIS break junctions and SIN junctions 
were prepared on freshly cleaved surfaces by a point contact technique 
with a Au tip at 4.2 K to reduce any surface contaminations.  First 
SIN junctions are obtained, followed by SIS junctions formed by 
breaking off a piece of the Bi2212 sample which attaches to the Au 
tip \cite{OzyuzerPRB}. $I-V$ and $dI/dV-V$ characteristics are recorded 
using the conventional lock-in technique \cite{Ozyuzer}. 

\begin{figure}[tb]
\vskip.15in
\centerline{
\begin{minipage}{\linewidth}
\centerline{\epsfxsize=\linewidth 
\epsfbox{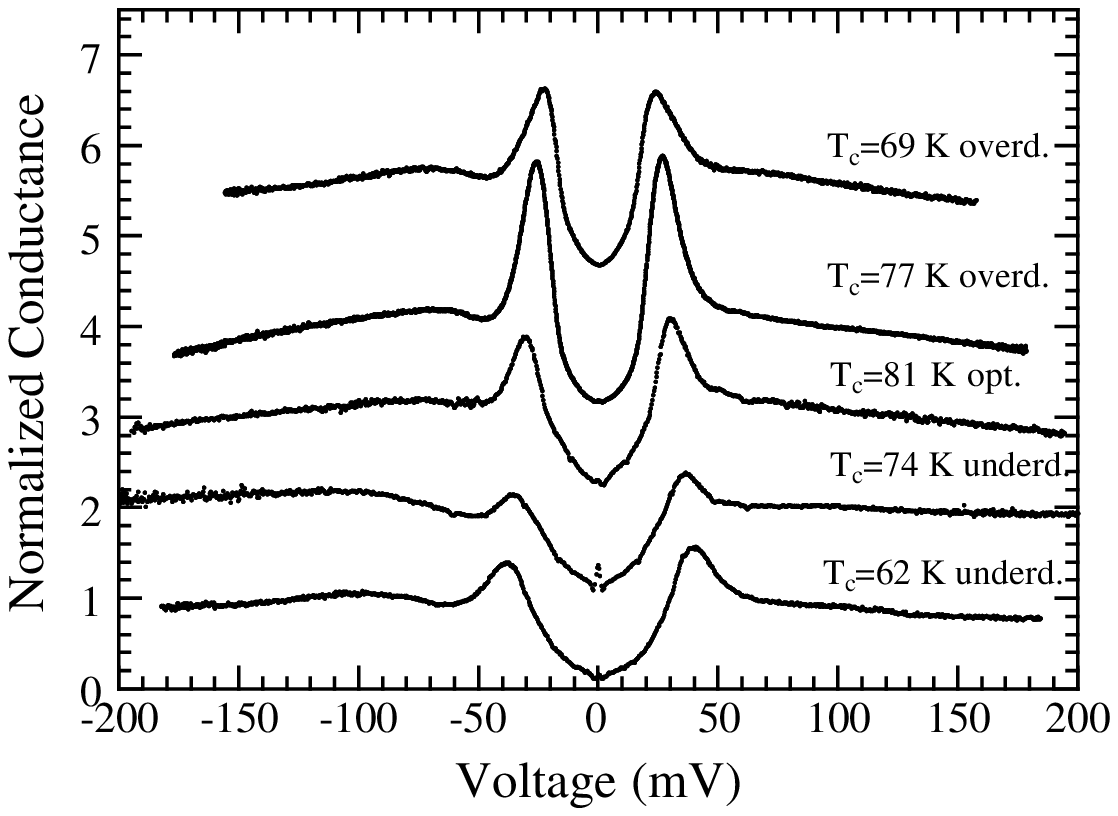}} 
%\vskip.15in 
\caption{Tunneling conductances of five SIN junctions 
on five different Ca-rich Bi2212 crystals, each with a different oxygen 
concentration.  Each spectra is normalized by a constant which is 
obtained from the high bias conductance value.  The data are taken at 4.2 
K and shifted for clarity.}
\label{1}
\end{minipage}}
\end{figure}  
 
\section{Results and Discussion}

Fig.\ 1 shows the tunneling conductances of five SIN junctions at 4.2 K 
obtained from five different crystals with $T_{c}$=69 and 77 K 
overdoped, $T_{c}$=81 K optimally doped, $T_{c}$=74 K and $T_{c}$=62 K 
underdoped.  Here and hereafter in figures the voltage is the sample 
voltage which corresponds to removal of electrons from the 
superconductor, in other words the negative bias is the occupied 
quasiparticle states in the DOS.  Each spectrum is normalized by a 
constant and shifted vertically for clarity.  Each spectrum exhibits 
quasiparticle peaks at $eV$$\sim$$\pm\Delta$, which indicates different 
energy gap magnitudes for different hole concentrations.  Here it is 
very important to notice that peculiar spectral features (dip and 
hump), which are always observed in careful tunneling and ARPES 
studies, \cite{Miyakawa}-\cite{Oda,OzyuzerPRB,OzyuzerEPL}-\cite{Pan2001}
exist over the entire doping range indicating close relation to 
superconductivity.  As the oxygen doping decreases from the overdoped 
region to the underdoped region, the energy gap magnitude increases as 
in nonsubstituted crystals of Bi2212.  Furthermore, the overall 
spectral features persist and move out to higher voltages as the gap 
increases.  Except for the particular overdoped $T_{c}$=77 K sample, 
there is a general decrease of the conductance peak height to 
background ratio with decreasing oxygen concentration, which is 
consistent with recent ARPES studies which show that the quasiparticle 
spectral weight is more broadened in the underdoped 
regime \cite{Campuzano,Feng}. 

A reproducibly obtained characteristic in tunneling studies of 
stoichiometric 
Bi2212 is a conductance peak height asymmetry that depends on the hole 
concentration \cite{Miyakawa,DeWilde,Renner96}. In the overdoped regime the peak 
in the occupied states is considerably higher than the one in the unoccupied 
states and this cannot be removed from the background by normalization. 
In the underdoped regime, the asymmetry is just the opposite and the 
crossover occurs near optimal doping.  Recent models of asymmetry in 
tunneling studies of Bi2212 \cite{Hirsh,zikri} have suggested that this 
might be an intrinsic effect.  However, another tunneling study of Bi2212 
by scanning tunneling microscopy/spectroscopy (STM/STS) shows a more 
consistent asymmetry between overdoped and underdoped 
Bi2212 \cite{Pan2001}. Since the tunneling conductances obtained by 
STM/STS may be susceptible to an energy dependent background, such 
effects need to be considered as well.  On the other hand, the 
quasiparticle spectral peaks in Ca-rich Bi2212 in Fig.\ 1 show an 
asymmetry whereby the peaks at the unoccupied side of DOS are higher.  
Only the most overdoped sample exhibits the more standard asymmetry.  
If there is a shift of the asymmetry, it is occurring somewhere in the 
overdoped region.  Furthermore, for a given doping level (e.g.  
optimal $T_{c}$) the Ca-rich spectra appear more broadened than the 
stoichiometric spectra.  This may be a consequence of the increased 
disorder from the excess Ca.

\begin{figure}[tb]
\vskip.03in 
\centerline{
\begin{minipage}{\linewidth}
\centerline{\epsfxsize=\linewidth 
\epsfbox{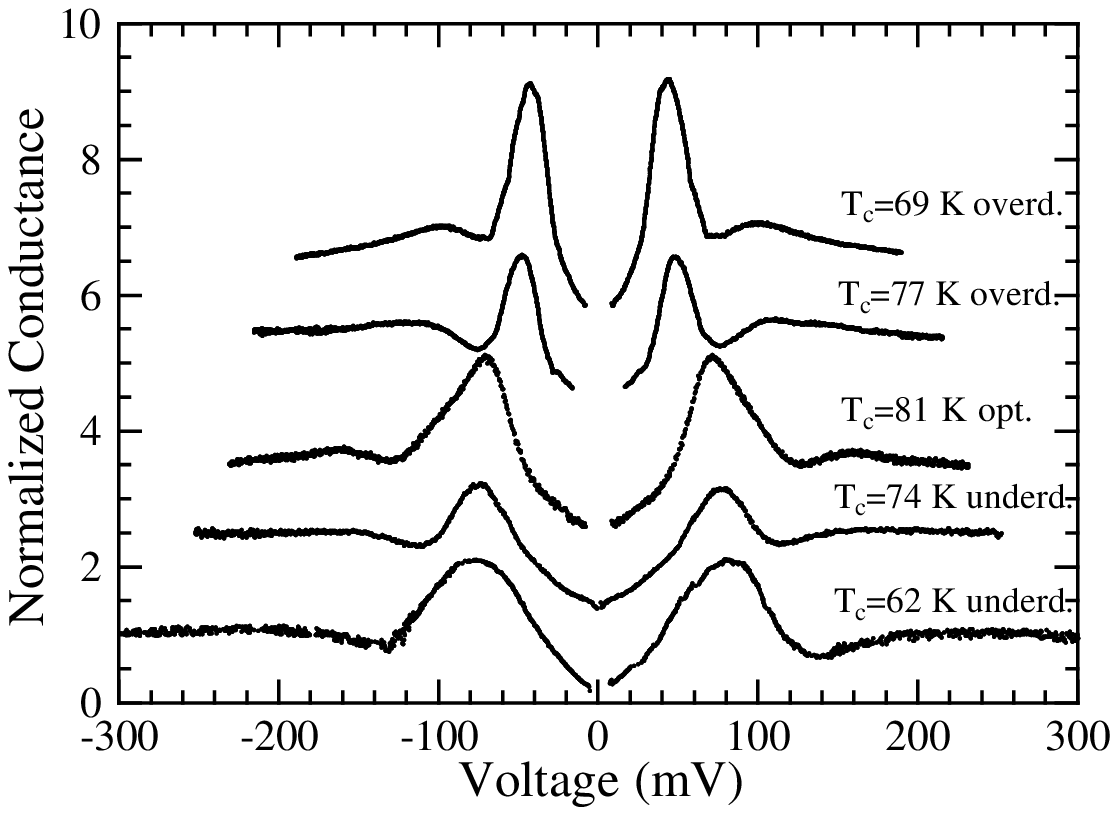}} 
\caption{Tunneling conductances of five representative SIS junctions 
on five different Ca-rich Bi2212 crystals, each with a different oxygen 
concentration.  The data are taken at 4.2 K.  The spectra (except 
$T_{c}$=62 K underdoped) are shifted and Josephson current peaks are 
removed from the data for clarity.}
\label{2}
\end{minipage}}
\end{figure}
         
It is useful to obtain SIS junctions not only as a consistency check but because the reduced 
smearing effects in SIS junctions allow a direct measure of the energy gap 
from the quasiparticle 
peak positions ($\pm$2$\Delta$). Also, the high-bias spectral features are 
amplified in the SIS geometry.  Fig.\ 2 shows normalized tunneling 
conductance-voltage characteristics of five SIS junctions from five 
different hole concentrations ($T_{c}$=69 and 77 K overdoped, 
$T_{c}$=81 K optimally doped, $T_{c}$=74 and 62 K underdoped).  These 
were measured at 4.2 K, normalized by a constant (at 200 mV) and 
shifted for clarity.  The Josephson current peaks are erased for 
clarity and will not be discussed in this paper.  Similar to Fig.\ 1 
which shows SIN junctions, Fig.\ 2 also reveals an increasing energy 
gap magnitude with decreasing hole concentration for SIS junctions.  
At the same time, more pronounced dip and hump structures also shift 
towards higher energies with underdoping.  Note that the peaks 
correspond to $\pm$2$\Delta$ and therefore the characteristic energies 
of the dip and hump are shifted by an additional $\Delta$.  
It has been clearly shown that the SIN data can reproduce the SIS 
spectrum by convoluting it by itself \cite{Miyakawa}. Some of the SIS 
spectra exhibit rather large peak widths which may be an indication of 
an inhomogeneous gap distribution within the junction area.

\begin{figure}[tb]
\centerline{
\begin{minipage}{\linewidth}
\centerline{\epsfxsize=\linewidth 
\epsfbox{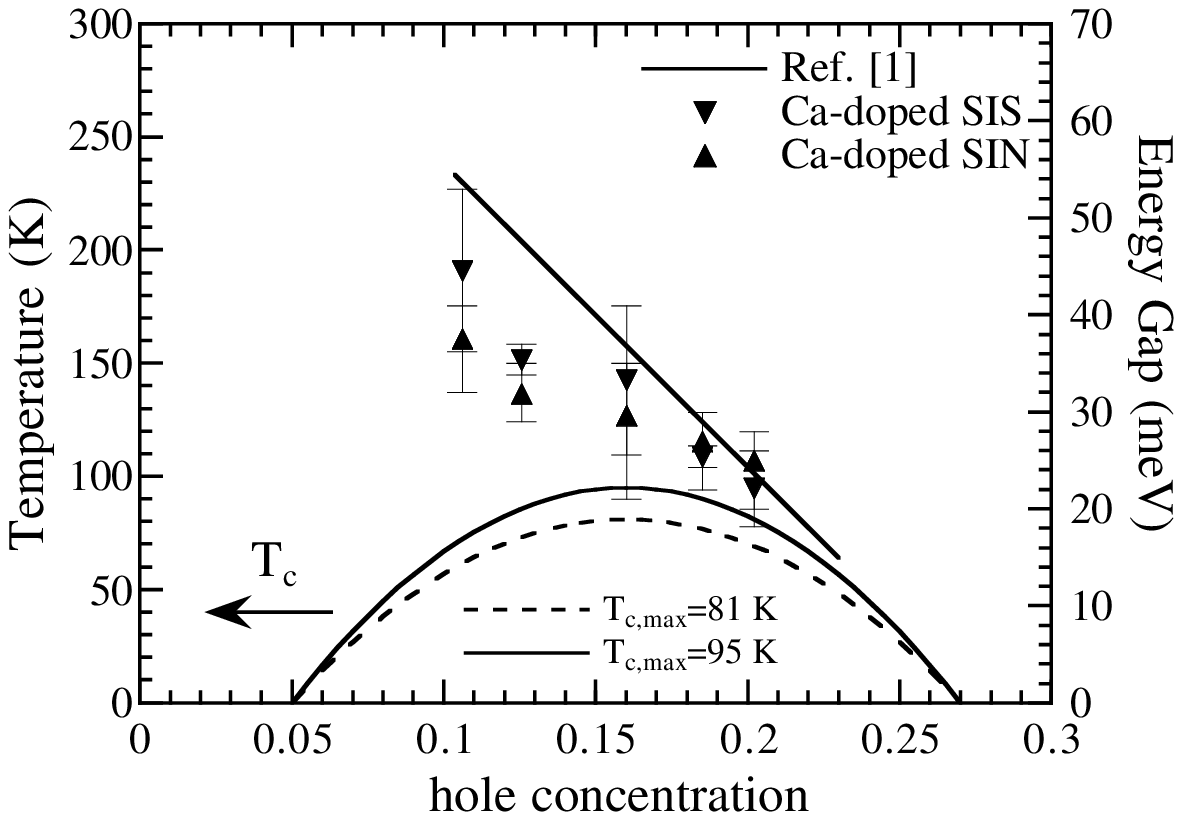}} 
\vskip.15in 
\caption{The phase diagram of the Bi2212 system.  The solid and 
dashed curves correspond to $T_{c}$ of the stoichiometric and Ca-rich Bi2212 
respectively using the emprical formula 
$T_{c}$/$T_{c,max}$=1-82.6($p$-0.16)$^{2}$.  The solid line is the 
line fit to Ref.\ [1].  The triangles are the present data which are 
averages of energy gap values obtained from SIN and SIS tunneling 
experiments (see legend).}
\label{3}
\end{minipage}}
\end{figure}

In Fig.\ 3 are shown the measured gap 
values vs. hole concentration in Ca-rich Bi2212.  The solid and 
dashed curve are the $T_{c}$ for stoichiometric Bi2212 and Ca-rich 
Bi2212 respectively, where the maximum critical temperature, 
$T_{c,max}$, 
is 95 K for stoichiometric Bi2212 and 81 K for Ca-rich Bi2212.  The 
hole concentrations, $p$, are estimated from the empirical formula 
$T_{c}$/$T_{c,max}$=1-82.6($p$-0.16)$^{2}$.  The solid line is the 
linear fit 
to our previously published data \cite{Miyakawa}. 
The gap magnitudes are estimated directly from the quasiparticle peaks without 
any fit to $d$-wave BCS theory.  Contrary to our 
previous 
observations on stoichiometric Bi2212 which show similar gap 
magnitudes for SIN and SIS junctions, in Ca-rich Bi2212 we observed 
not only a large scattering of energy gap magnitudes, but also an 
accumulation of averaged gaps at different energy locations for two 
different types of junction geometry, SIN and SIS.  Each hole 
concentration depicts at least 10 SIN and SIS junctions to get 
reasonable statistical distribution of the energy gap magnitudes.  
The gap 
magnitude average of SIS junctions is larger than the gap magnitude 
average of SIN junctions in the underdoped range.  However, over the 
wide doping range which is examined, the error bars of both junction 
geometries lie top of each other.  Our use of the peak voltage in SIN 
junctions leads to an overestimate of the gap, which makes the 
discrepancy between SIS and SIN gap values a bit larger.  The origin of 
this discrepancy is not known but perhaps the increased defect density in the 
Ca-rich Bi2212 allows a higher diffusion rate for oxygen.  Since the 
equilibrium oxygen concentration in air leads to slight overdoping, this 
tendency would give a smaller gap on the surface compared to the bulk, the 
latter being probed by SIS break junctions.

While the average gap 
magnitude is smaller than for stoichiometric Bi2212, at least part of this 
is due to the generally lower $T_{c}$ value for any hole concentration in 
Ca-rich Bi2212.  Note however that the overall 
doping dependence shows the same trend, that is, the gap increases in the 
underdoped region even as $T_{c}$ decreases.

\begin{figure}[tb]
%\vskip-.02in 
\centerline{
\begin{minipage}{\linewidth}
\centerline{\epsfxsize=\linewidth 
\epsfbox{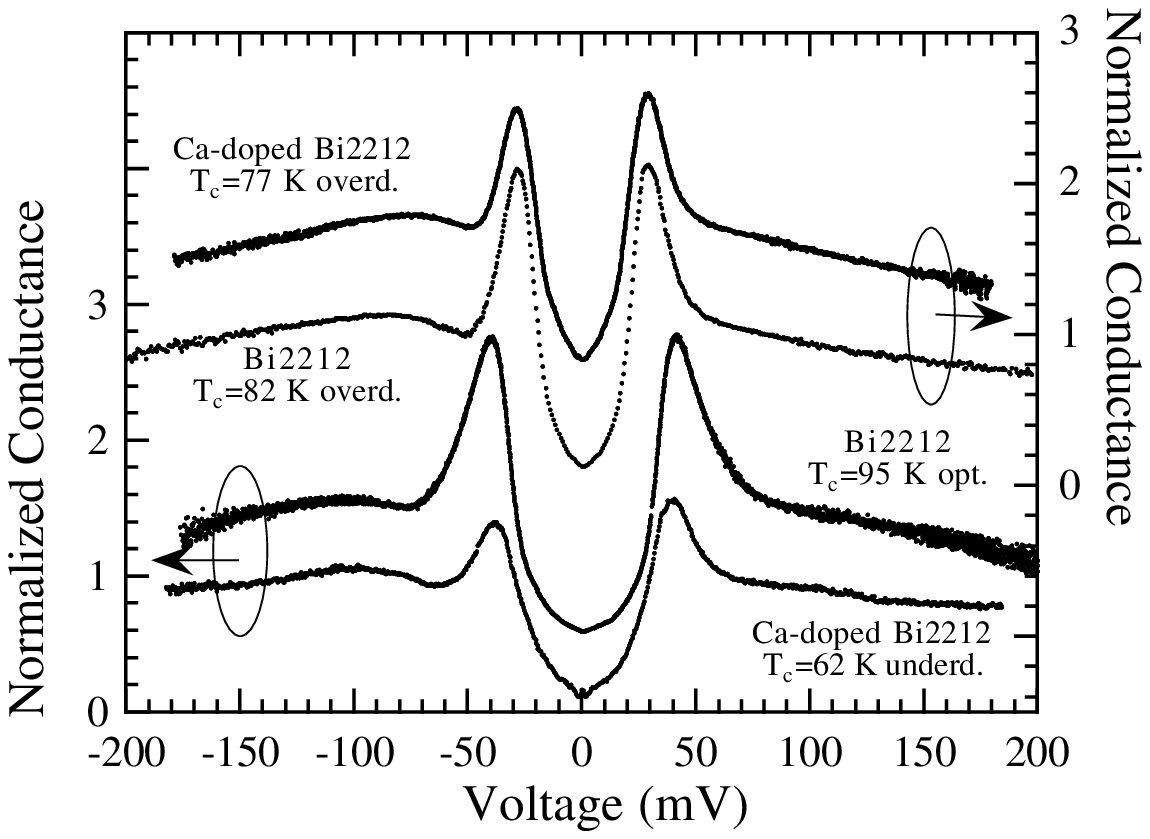}} \vskip.11in 
\caption{Comparison of stoichiometric and Ca-rich Bi2212 SIN data.  
The upper two spectra are overdoped crystals with similar gap values 
and $T_{c}$'s.  While the lower two spectras have similar gaps, they 
have different $T_{c}$'s.}
\label{4}
\end{minipage}}
\end{figure}

In order to understand the spectral features of stoichiometric and Ca-rich Bi2212, 
we compare spectra that have the same gap magnitude. For example, we plotted two 
overdoped SIN junctions which have similar gap 
magnitudes in the top part of Fig.\ 4.  Note that the $T_{c}$ values 
are not that different.  Both tunneling conductances are normalized by 
a constant and Ca-doped Bi2212 is shifted by 0.7 units.  This figure 
shows that the tunneling conductances are nearly identical, showing 
the same $d$-wave like subgap conductance, sharp quasiparticle peaks, 
and dip/hump features at similar energies.  The bottom part of Fig.\ 4 
shows optimally doped $T_{c}$=95 K stoichiometric Bi2212 compared with 
underdoped $T_{c}$=62 K Ca-rich Bi2212, both of which have gap values 
near 40 meV.  While the hump energies are nearly the same, the 
location of the dip feature is considerably different in the two 
spectra as are the critical temperatures.  The latter effect is 
consistent with previous observations that the dip location follows 
the neutron resonance mode energy which scales with $T_{c}$ and not 
the gap energy \cite{Zasadzinski}. The smaller $T_{c}$ value of the 
Ca-rich sample would thereby imply a smaller resonance mode energy 
which would put the dip location closer to the gap edge.  At present 
these results are qualitative and await further quantitative analysis.

In summary, we have prepared Ca-rich Bi2212 crystals and varied the doping 
via the oxygen concentration.  The optimum $T_{c}$ value of the Ca-rich 
samples is 81 K, significantly smaller than the optimum value of 95 K in 
stoichiometric Bi2212.  The key result is that the trend of the energy gap 
vs. doping is the same in both types of Bi2212.  This suggests that this 
trend is a universal feature of high $T_{c}$ cuprates.  Preliminary 
analysis of the tunneling spectra suggests that the hump feature scales 
directly with the energy gap but that the dip feature is tied more closely 
to the bulk $T_{c}$.  Such a behavior is consistent with previous 
investigations which suggest the dip feature is linked to the 
resonance mode found in neutron scattering.

% insert where needed to balance the two columns on the last page
%\newpage
 
\end{document}